\documentclass[prl,twocolumn,superscriptaddress,showpacs]{revtex4} 
\usepackage{bm}

\usepackage{times}
\usepackage{amsmath}
\usepackage{amsthm}
\usepackage{amssymb}
\usepackage{amsbsy}
\usepackage{graphicx}
\usepackage{pstricks}
\usepackage{wasysym}

\begin{document}

\newcommand{\smallrech}{ \; \pspicture(0.2,0.1)
\psset{linewidth=0.025,linestyle=solid}
\psline[](0.1,0)(0.1,0.1)
\pspolygon[](0,0)(0.2,0)(0.2,0.1)(0,0.1)
\endpspicture\;}
\newcommand{\orients}{ \; \pspicture(0.01,0.01) \psarc{->}(-0.01,0.12){0.1}{10}{350} \endpspicture\;}

\newcommand{\smallrecv}{ \; \pspicture(0.1,0.2)
\psset{linewidth=0.025,linestyle=solid}
\psline[](0,0.1)(0.1,0.1)
\pspolygon[](0,0)(0,0.2)(0.1,0.2)(0.1,0.0)
\endpspicture\;}

\newcommand{\emptya}{  \; \pspicture(0.6,0.2)
\psdots[linecolor=black,dotsize=.10]  (0.15,0)(0.15,0.3)  (0.6,0.15)
\psset{linewidth=0.03,linestyle=dotted}
\psline[](0.3,0)(0.3,0.3)
\pspolygon[](0,0)(0.6,0)(0.6,0.3)(0,0.3)
\psset{linewidth=0.05,linestyle=solid}
\psline[linecolor=black](0,0)(0.3,0)
\psline[linecolor=black](0,0.3)(0.3,0.3)
\psline[linecolor=black](0.6,0)(0.6,0.3)
\endpspicture \;}

\newcommand{\emptyb}{ \; \pspicture(0.6,0.2)
\psdots[linecolor=black,dotsize=.10]  (0.45,0)(0.45,0.3)(0,0.15)
\psset{linewidth=0.03,linestyle=dotted}
\psline[](0.3,0)(0.3,0.3)
\pspolygon[](0,0)(0.6,0)(0.6,0.3)(0,0.3)
\psset{linewidth=0.05,linestyle=solid}
\psline[linecolor=black](0.3,0)(0.6,0)
\psline[linecolor=black](0.3,0.3)(0.6,0.3)
\psline[linecolor=black](0,0)(0,0.3)
\endpspicture \;}

\newcommand{\fulla}{ \; \pspicture(0.6,0.2)
\psdots[linecolor=black,dotsize=.10]   (0.15,0)(0.15,0.3)  (0.6,0.15)(0.3,0.15)
\psset{linewidth=0.03,linestyle=dotted}
\psline[](0.3,0)(0.3,0.3)
\pspolygon[](0,0)(0.6,0)(0.6,0.3)(0,0.3)
\psset{linewidth=0.05,linestyle=solid}
\psline[linecolor=black](0,0)(0.3,0)(0.3,0.3)(0,0.3)
\psline[linecolor=black](0.6,0)(0.6,0.3)
\endpspicture\;}

\newcommand{\fullb}{ \; \pspicture(0.6,0.2)
\psdots[linecolor=black,dotsize=.10]  (0.45,0)(0.45,0.3)  (0,0.15)(0.3,0.15)
\psset{linewidth=0.03,linestyle=dotted}
\psline[](0.3,0)(0.3,0.3)
\pspolygon[](0,0)(0.6,0)(0.6,0.3)(0,0.3)
\psset{linewidth=0.05,linestyle=solid}
\psline[linecolor=black](0.6,0)(0.3,0)(0.3,0.3)(0.6,0.3)
\psline[linecolor=black](0,0)(0,0.3)
\endpspicture \;}

\newcommand{\hexaa}{\;\pspicture(0,0.1)(0.35,0.6)\psset{unit=0.75cm}
\psset{linewidth=0.03,linestyle=dotted}
\pspolygon(0,0.15)(0,0.45)(0.2598,0.6)(0.5196,0.45)(0.5196,0.15)(0.2598,0)
\psset{linewidth=0.08,linestyle=solid}
\psline(0,0.15)(0,0.45)
\psline(0.2598,0.6)(0.5196,0.45)
\psline(0.5196,0.15)(0.2598,0)
\psdots[linecolor=gray,dotsize=.20](0,0.30)
\psdots[dotstyle=triangle*,linecolor=gray,dotsize=.20](0.3897,0.525)
\psdots[linecolor=gray,dotsize=.20,dotstyle=square*](0.3897,0.075)
\endpspicture\;}

\newcommand{\hexab}{\;
\pspicture(0,0.1)(0.35,0.6)
\psset{unit=0.75cm}
\psset{linewidth=0.03,linestyle=dotted}
\pspolygon(0,0.15)(0,0.45)(0.2598,0.6)(0.5196,0.45)(0.5196,0.15)(0.2598,0)
\psset{linewidth=0.08,linestyle=solid}
\psline(0,0.15)(0,0.45)
\psline(0.2598,0.6)(0.5196,0.45)
\psline(0.5196,0.15)(0.2598,0)
\psdots[dotstyle=square*,linecolor=gray,dotsize=.20](0,0.30)
\psdots[linecolor=gray,dotsize=.20](0.3897,0.525)
\psdots[dotstyle=triangle*,linecolor=gray,dotsize=.20](0.3897,0.075)
\endpspicture\;}

\newcommand{\hexb}{\;
\pspicture(0,0.1)(0.35,0.6)
\psset{unit=0.75cm}
\psset {linewidth=0.03,linestyle=dotted}
\pspolygon[](0,0.15)(0,0.45)(0.2598,0.6)(0.5196,0.45)(0.5196,0.15)(0.2598,0)
\psset{linewidth=0.08,linestyle=solid}
\psline(0.2598,0.6)(0,0.45)
\psline(0.5196,0.15)(0.5196,0.45)
\psline(0.2598,0)(0,0.15)
\psdots[linecolor=gray,dotsize=.20](0.1299,0.525)
\psdots[dotstyle=triangle*,linecolor=gray,dotsize=.20](0.525,0.3)
\psdots[linecolor=gray,dotsize=.20,dotstyle=square*](0.1299,0.075)
\endpspicture\;}

\newcommand{\smallhexh}{ \;
\pspicture(0,0.1)(0.2,0.3)
\psset{linewidth=0.03,linestyle=solid}
\pspolygon[](0,0.0775)(0,0.225)(0.124,0.3)(0.255,0.225)(0.255,0.0775)(0.124,0)
\endpspicture
\;}


\title{Strings in strongly correlated electron systems}

\author{Peter Fulde}
\address{Max-Planck-Institut f{\"u}r Physik komplexer Systeme, 01187 Dresden, Germany}
\address{Asia Pacific Center for Theoretical Physics, Pohang, Korea}
\author{Frank Pollmann}
\address{Max-Planck-Institut f{\"u}r Physik komplexer Systeme, 01187 Dresden, Germany}

\date{\today}

\begin{abstract}
It is shown that strongly correlated electrons on frustrated lattices like
pyrochlore, checkerboard or kagom\`e lattice can lead to the appearance of
closed and open strings. They are resulting from nonlocal subsidiary conditions
which propagating strongly correlated electrons require. The dynamics of the
strings is discussed and a number of their properties are pointed out. Some of
them are reminiscent of particle physics.\\
\end{abstract}

\maketitle

\label{Sect:Introduction}

Strongly correlated electrons are a subject of intense investigation by many
theory groups. Part of that big interest is due to the fact that the
corresponding materials have often physical properties which make them
attractive for applications. The high-T$_c$ superconducting cuprates
\cite{Kuzmany,Bednorz,Plakida95} are perhaps the most prominent
examples. But also materials with a giant or collosal magnetoresistance like the
manganites have strongly correlated conduction electrons
\cite{Dagotto,Imada98,Kudasov03}. Other common examples include 4$f$
and 5$f$ electron systems with heavy quasiparticles at low temperatures
\cite{Stewart,Norman}, although here the practical use of those
materials is less obvious. 

Distinct from possible practical uses strongly correlated electron systems are
challenging many-body systems which are of basic interest, because techniques
for dealing with them must be developed which may apply also to other strong
coupling theories. We speak of strongly correlated electrons when the mutual
Coulomb repulsion of the particles influences their time evolution more
strongly than the kinetic energy gain due to delocalization. Hence the strong
coupling character of the many-body problem is obvious. The strong coupling
limit of an electron system depends on the type of lattice on which the
particles are moving as well as on the filling factor, i.e., on the ratio $n$
of electron number $L$ to number of lattice sites $N$. In the limit in which
particle hopping between sites is neglected, the ground state is usually
degenerate. Of special interest are geometrically frustrated lattices like the
pyrochlore lattice, its two-dimensional version the checkerboard lattice or the
kagom\`e lattice. Here the ground state is macroscopically degenerate for most
filling factors $n$. Adding the dynamics to such a system reduces the
degeneracy to a small number as examples discussed below will show.

The motion of strongly correlated particles underlies a large number of
nonlocal subsidiary conditions. They ensure that the repulsive particle
interactions which dominate the dynamics remain as small as possible. It will
be shown that particle propagation (worldlines) in the presence of those strong
nonlocal subsidiary conditions is equivalent to propagation of closed (loops)
or open strings, i.e., worldsheets with trivial subsidiary conditions. For
example, fully spin-polarized electrons (or spinless fermions) with strong
nearest-neighbor repulsion on a pyrochlore or checkerboard lattice at half
filling lead to a complete loop covering of the lattice when the ground state
is studied. The loops are obtained by connecting the occupied sites of the
lattice. Time evolution of loops, i.e., worldsheets is a phenomenon akin to
string theory. For a textbook on the theory see, e.g., Ref.~\cite{ZwiebachBook}.  

In the following we want to discuss the properties of closed and open strings
of some strongly correlated electron systems. We are dealing here with a
particularly simple example of a string theory which is numerically
solvable. Some of the features it contains are also met in elementary particle
physics. That strongly correlated electrons can be related to string theory
albeit of a different form as considered here has been realized before in
connection with the fractional quantum Hall effect. String theory realizations of the two-dimensional fractional quantum Hall fluid are found, e.g., in Refs.~\cite{Boyarsky,Bergmann}. The formation of strings in certain spin models has been described in Ref.~\cite{levin2005}.

\section{Strings on a checkerboard lattice}

\label{Sect:Strings}

\begin{figure}[thb]
\begin{center}
\includegraphics[height=57mm]{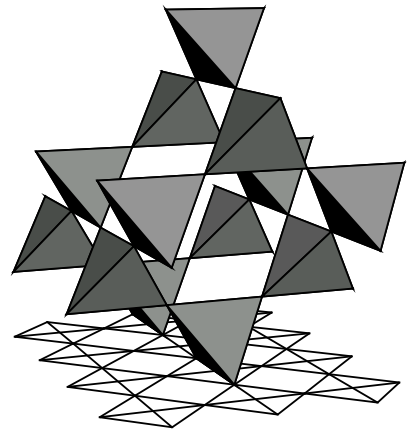}
\caption{Pyrochlore (above) and checkerboard lattice (below). The latter can be
considered as a projection of the former onto a plane (from Ref.
\cite{Moessner04a}).} 
\label{Fig1}
\end{center}
\end{figure}
~

\begin{figure}[thb]
 \begin{center}
      \begin{tabular}{cc}
	(a)~\includegraphics[width=3cm]{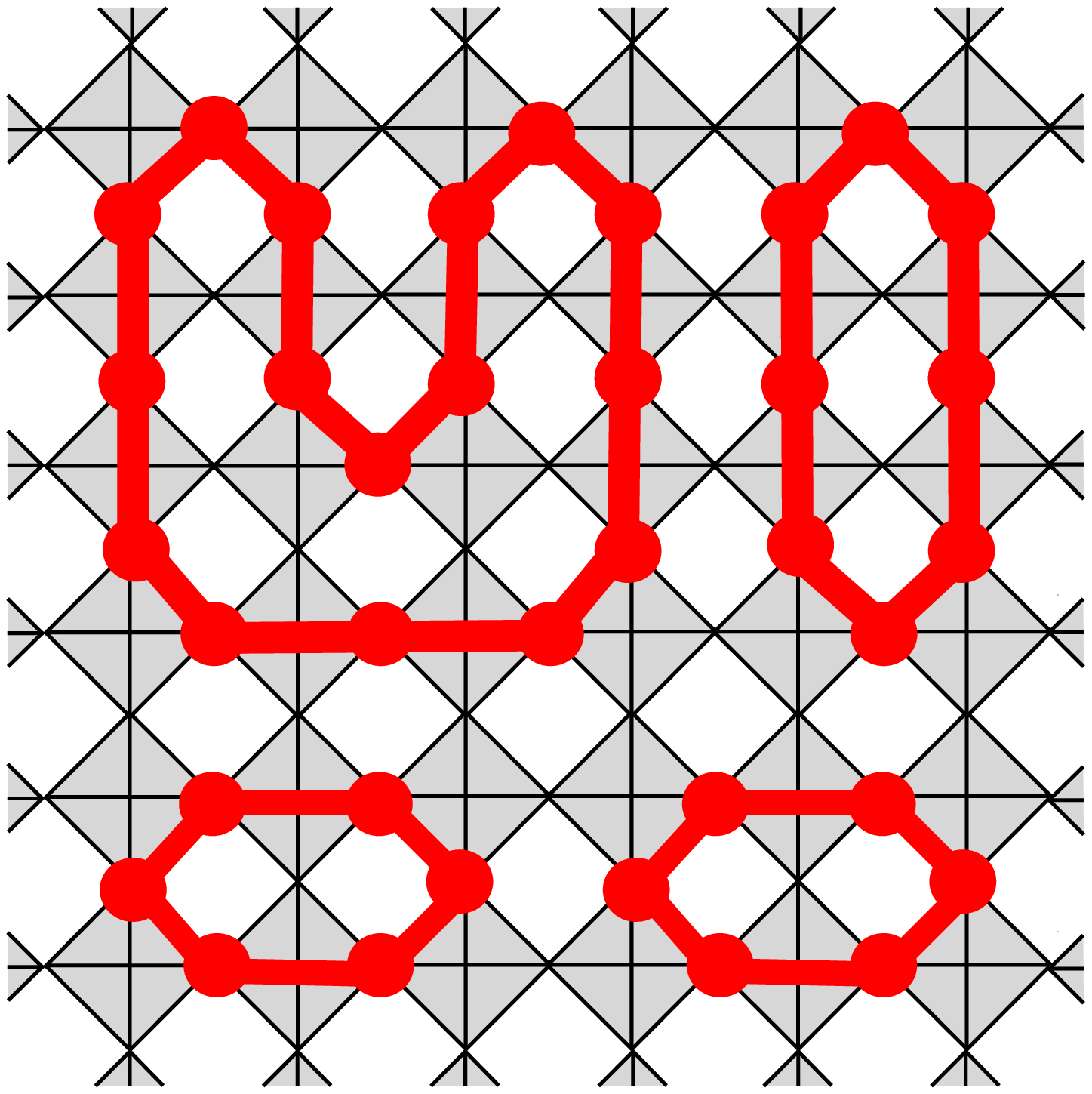}&
	(b)~\includegraphics[width=3cm]{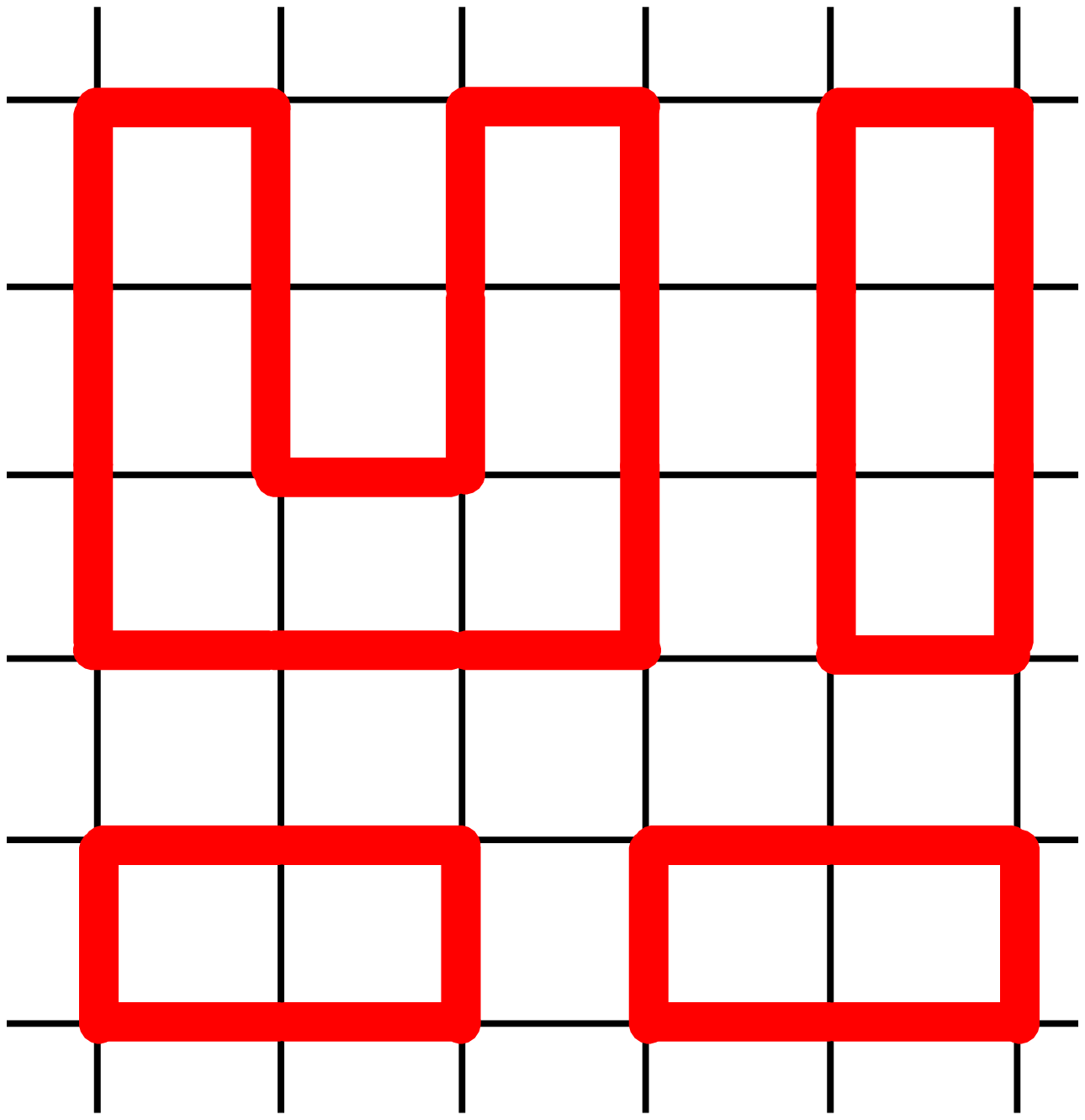}
      \end{tabular}
    \end{center}
    \caption{(a): Loop covering of the checkerboard plane for one of the ground    states. The loops are obtained by connecting sites occupied by spinless
    fermions. The case of half filling is considered.
(b): Loop covering in the medial, i.e., square lattice. The same ground-state
    configuration is shown as in (a).}
\label{Fig2}
\end{figure}
~
\begin{figure}[thb]
\includegraphics[height=57mm]{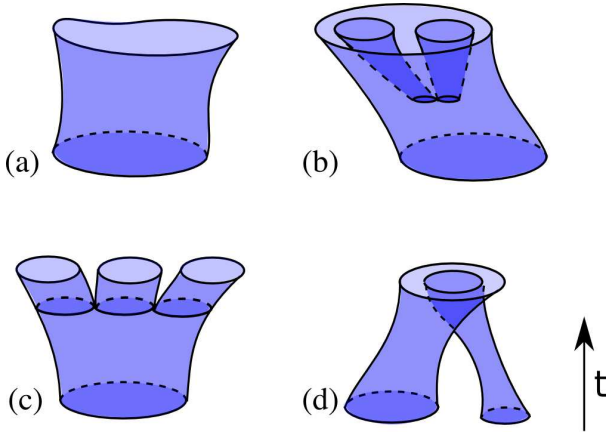}
\caption{Continuous representation of loop dynamics due to H$_{\rm eff}$.
(a): time evolution of loops due to B processes which conserve the topology.
(b-d): the same for A processes where H$_{\rm eff}$ induces three kinds of
  topological changes \cite{Pollmann06c}.}
\label{Fig3}
\end{figure}
~
\begin{figure}[thb]
 \begin{center}
      \begin{tabular}{cc}
	(a)\includegraphics[height=3cm]{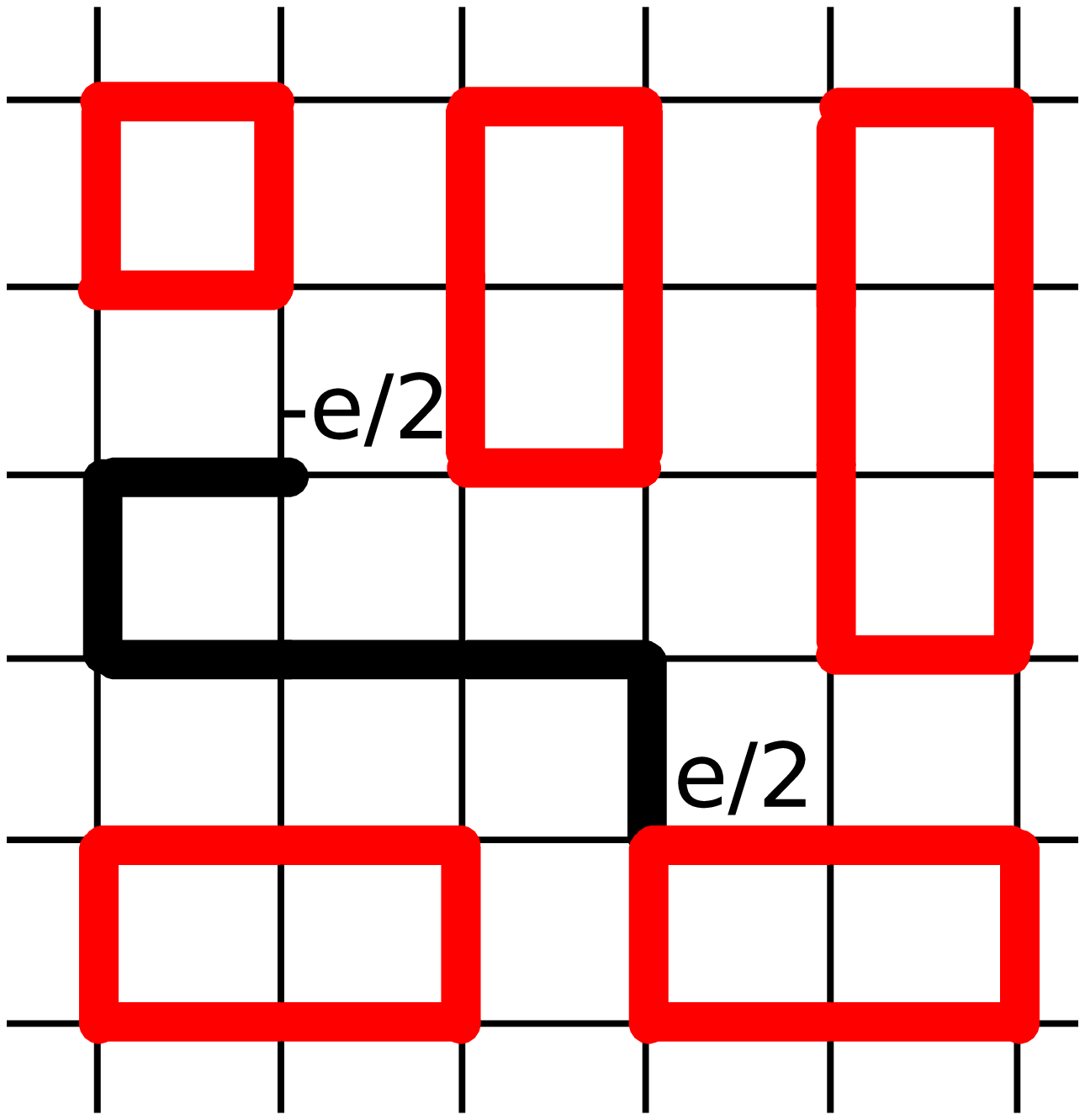}&
	(b)\includegraphics[height=3cm]{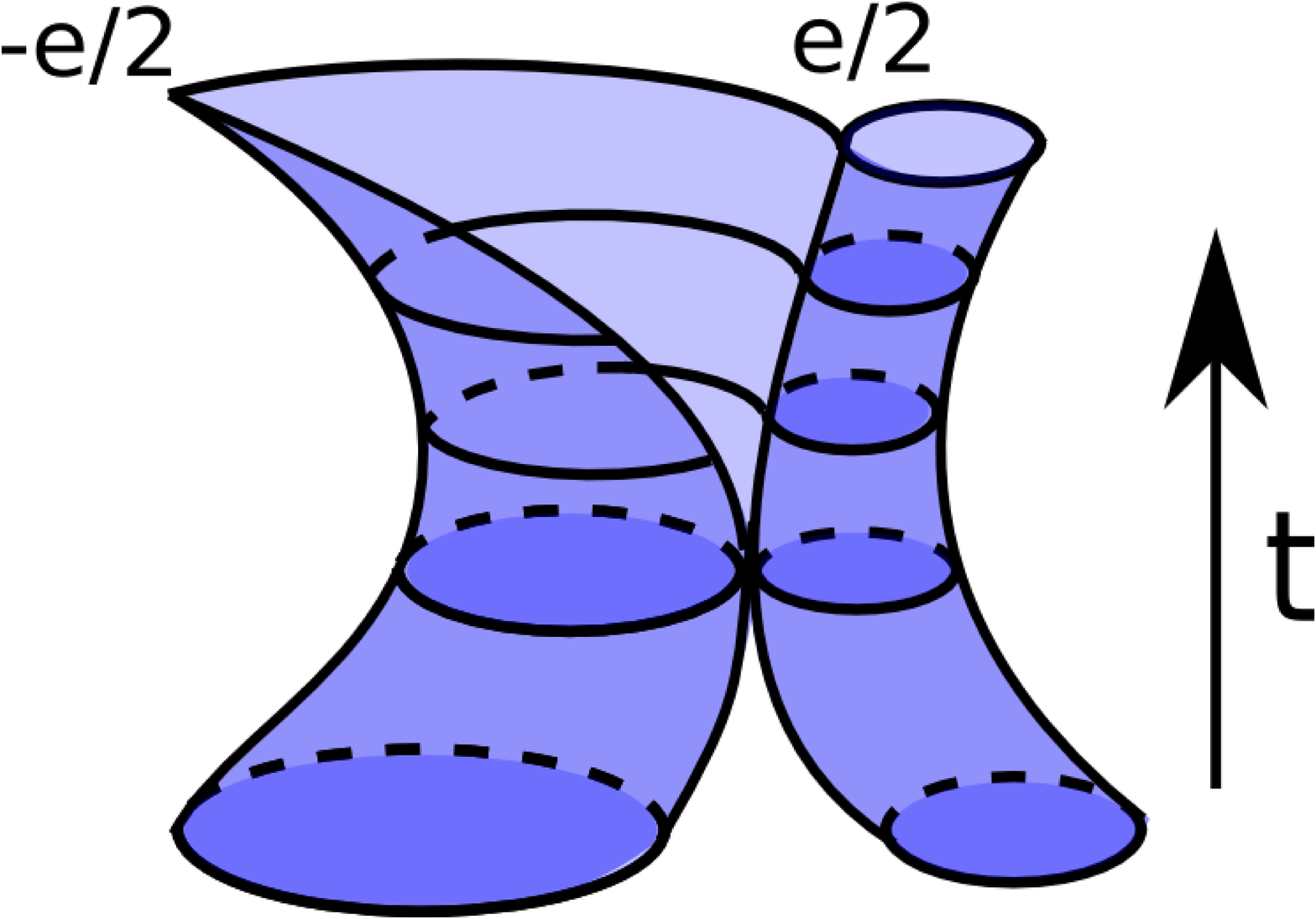}
      \end{tabular}
    \end{center}
    \caption{When an energy $\Delta E > V$ is added to the system a loop is
    broken and an open string is generated. Thereby one end of the string is
    touching a closed loop.
(a): medial lattice with the string shown by a black line.
(b): time evolution in a continuous representation.}
\label{Fig4}
\end{figure}
~
\begin{figure}[thb]
 \begin{center}
      \begin{tabular}{cc}
	(a)\includegraphics[height=3cm]{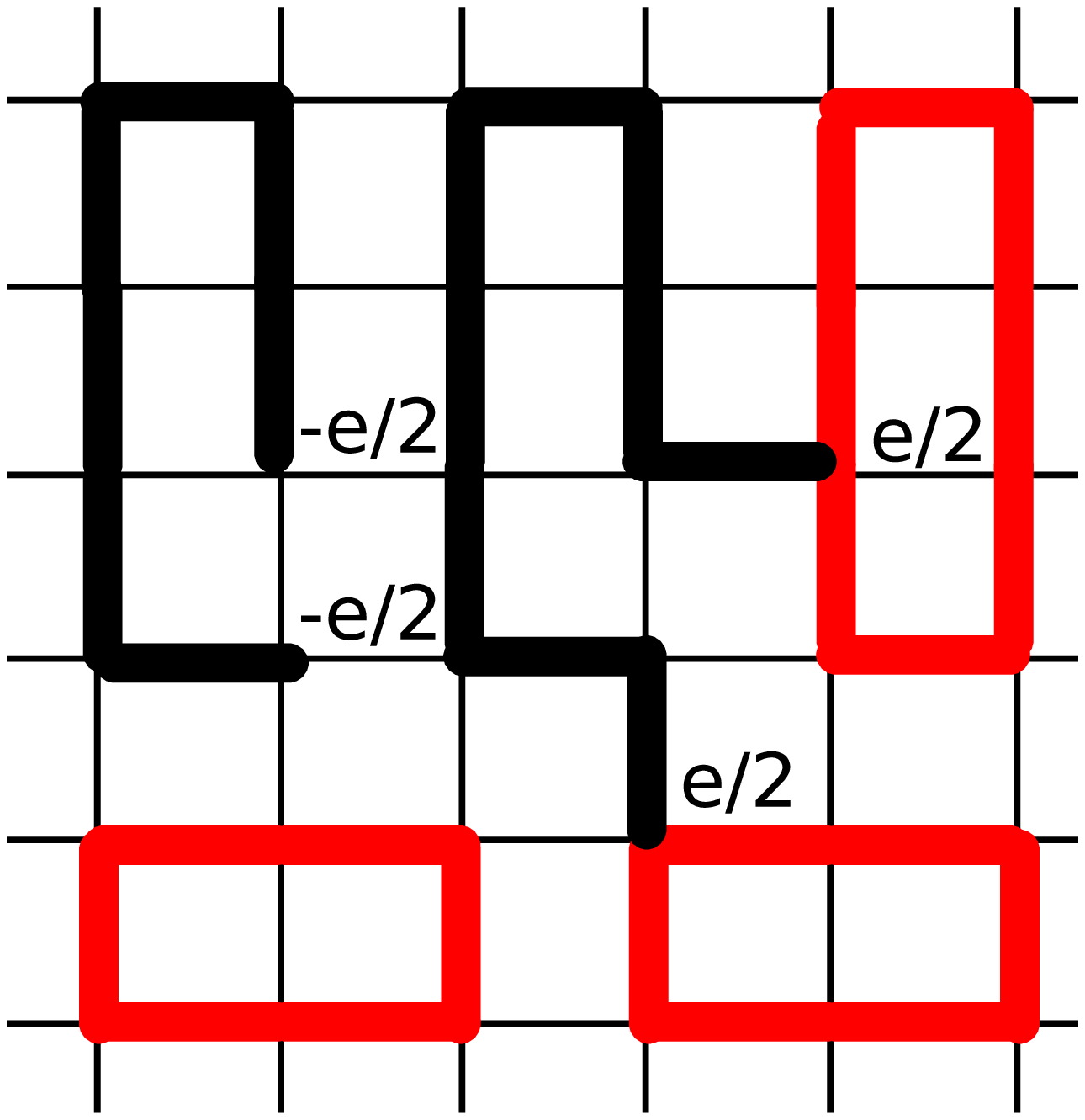}&
	(b)\includegraphics[height=2.7cm]{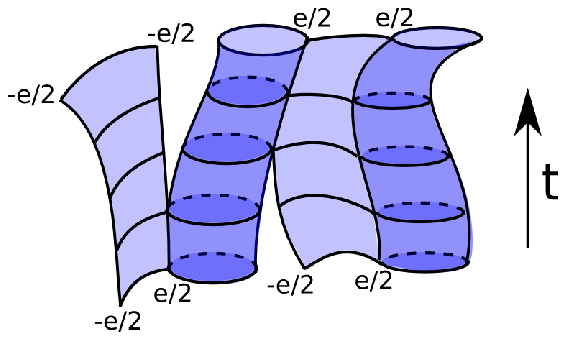}
      \end{tabular}
    \end{center}
    \caption{Configuration with two pairs $\left( \frac{e}{2}, \frac{e}{2}
    \right)$ and $\left( -\frac{e}{2}, -\frac{e}{2} \right)$ resulting from
    two particle-antiparticle pairs $\left( \frac{e}{2}, -\frac{e}{2} \right)$
    with fractional charges $\frac{e}{2}$ (or $-\frac{e}{2}$) of different
    colors (i.e., sublattice index).
(a): on the medial lattice;
(b): time evolution of two $\left( \frac{e}{2}, -\frac{e}{2} \right)$ pairs and
    formation of $e$ and $-e$ particles.} 
\label{Fig5}
\end{figure}
~
\begin{figure}[thb]
 \begin{center}
      \begin{tabular}{cc}
	(a)\includegraphics[height=4cm]{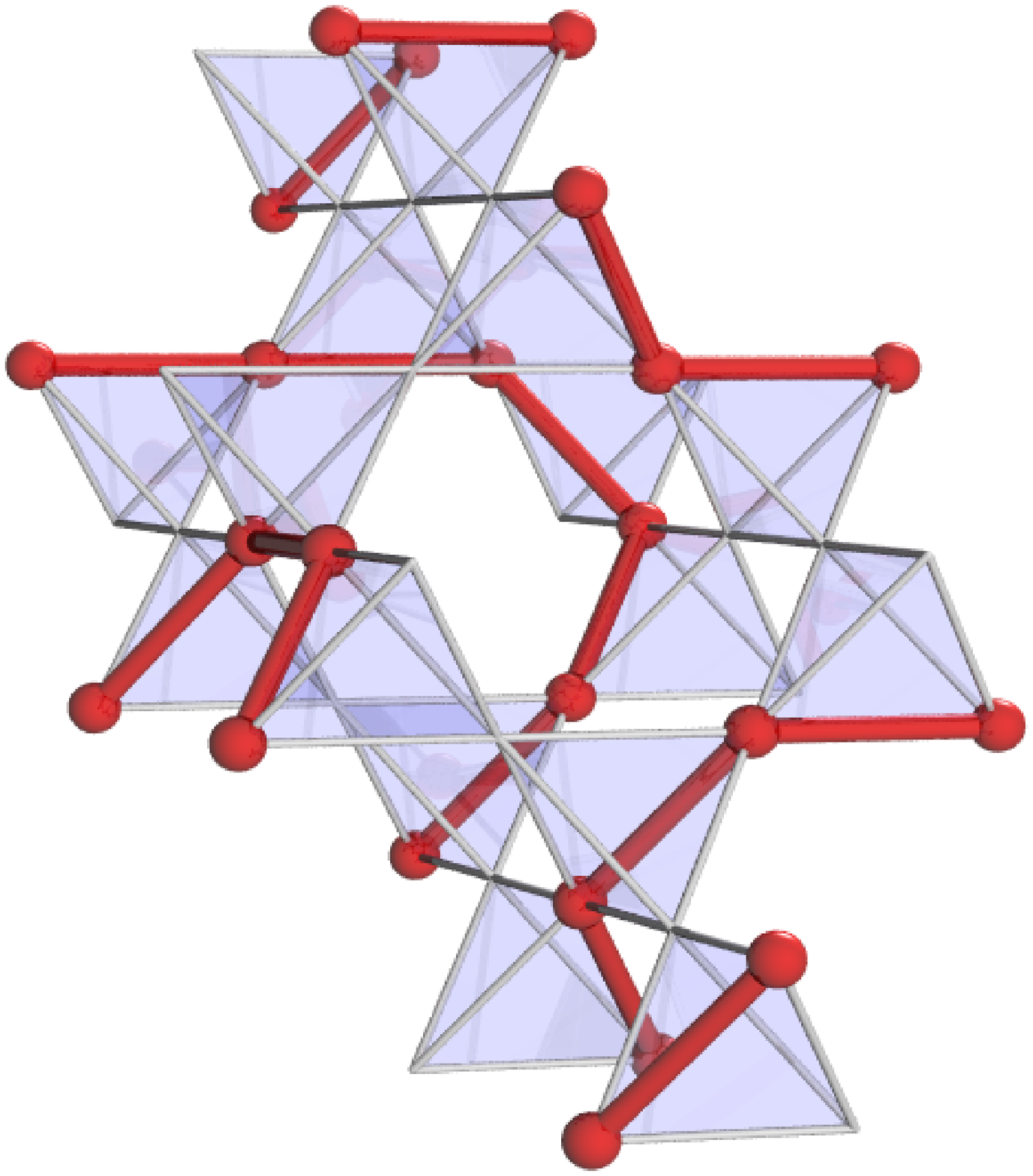}
	(b)\includegraphics[height=4cm]{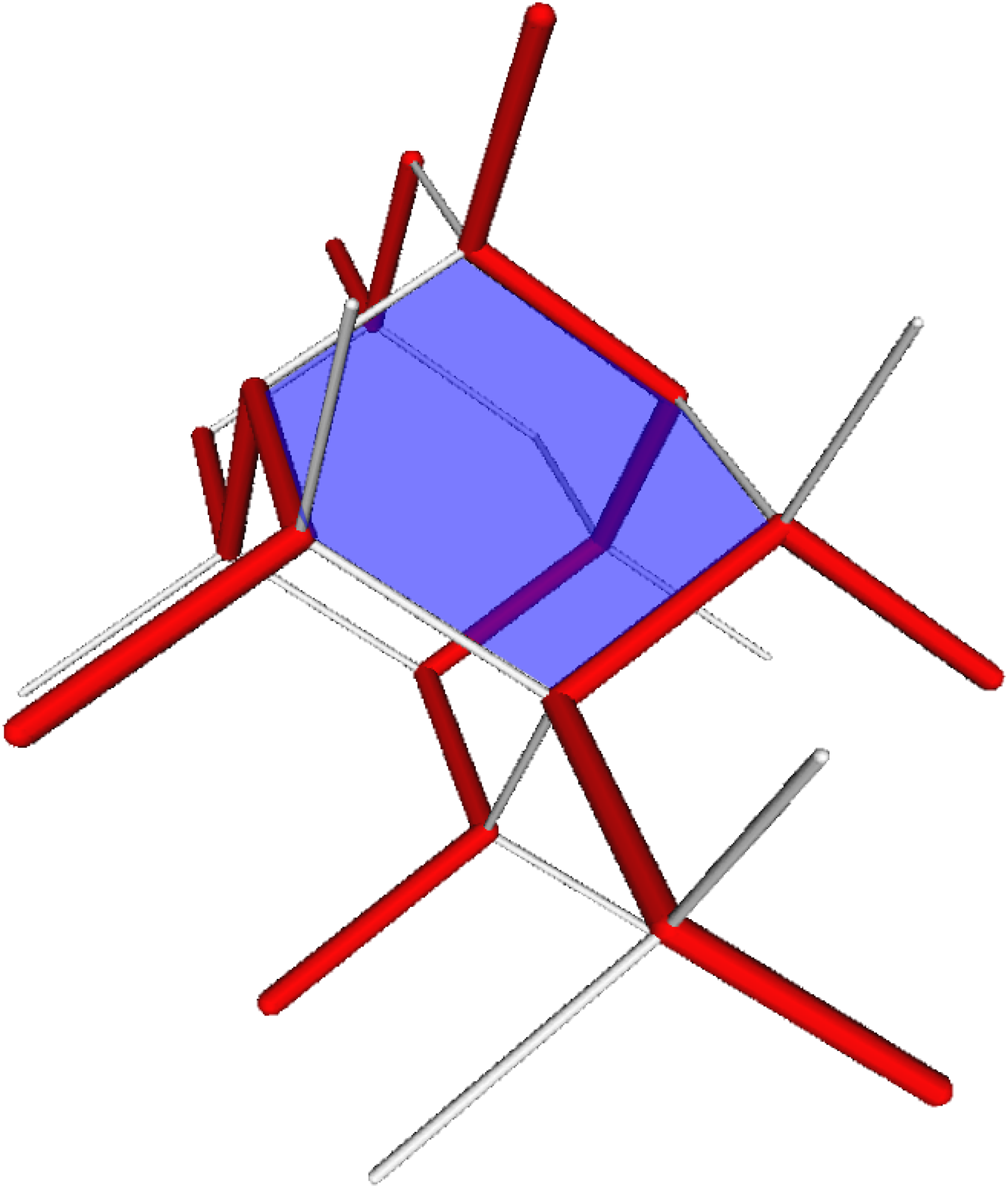}
      \end{tabular}
    \end{center}
    \caption{Example of a ground-state configuration of spinless fermions at
    half filling of a pyrochlore lattice (a) and shown for the medial, i.e.,
    diamond lattice (b). Here particles sit on links of the lattice structure.}
\label{Fig6}
\end{figure}


The checkerboard lattice can be considered as the two-dimensional version of a
pyrochlore lattice. A pyrochlore lattice consists of corner sharing
tetrahedra. Each lattice site has six nearest neighbors and the same holds true
for the checkerboard lattice (see Fig. \ref{Fig1}). 

We consider spinless (or fully spin polarized) fermions and half-filling $n =
\frac{1}{2}$, i.e., there are twice as many lattice sites $N$ than there are
particles. Thus we concentrate exclusively on charge degrees of freedom. The
Hamiltonian is 

\begin{equation}
H = -t \sum_{\langle ij \rangle} \left( c^+_i c_j + H.c. \right) + V
\sum_{\langle ij \rangle} n_i n_j
\label{eq1}
\end{equation} 

\noindent with $n_i = c^+_i c_i$ and $\langle ij \rangle$ denoting pairs of
neighboring sites. We assume strong coupling, i.e., $V \gg \mid t \mid$. In the
absence of hopping $(t = 0)$ the ground state is macroscopically
degenerate with degeneracy $N_{\rm deg} \simeq \left(
\frac{4}{3}\right)^{\frac{3}{4}N}$ \cite{Lieb67}. Each of the ground
states minimizes the repulsive energy $V$ by having two occupied and two empty
sites on each of the crisscrossed squares (''tetrahedra''). This is the
so-called ''tetrahedron rule'' \cite{Anderson56}. By connecting
nearest-neighbor occupied sites and applying periodic boundary conditions loops
are formed. The above subsidiary conditions lead to a macroscopic loop covering
of the plane for each of the ground states. For an example see Fig. \ref{Fig2}a
(each tetrahedron is touched by exactly one loop). 
 
In order to lift the high degeneracy of the ground state by the dynamics of the
system represented by $t$, it does not suffice to go to order $t^2/V$. In that
order dynamical processes contribute equally to all ground states. But the
extensive degeneracy is lifted in order $t^3/V^2$. To that order the
effective Hamiltonian becomes \cite{Runge04}

\begin{equation}
H_{\rm eff} = \frac{12 t^3}{V^2} \sum_{\smallhexh} c^+_{j_6} c^+_{j_4} c^+_{j_2}
c_{j_5} c_{j_3} c_{j_1}  
\label{eq2}
\end{equation} 

\noindent where the sum is over all hexagons. The subscripts $j_1, ..., j_6$
label the six sites of hexagon $j$. The effective Hamiltonian describes ring
hopping on hexagons. It connects different ground states. In order to discuss
the remaining degeneracy as well as the properties of strings it helps to go
over to the medial lattice which here is the square lattice. It is obtained from
the checkerboard lattice by connecting the centers of the crisscrossed
squares. In the medial lattice particles sit on links between sites. For illustration a ground-state configuration in the medial lattice is shown in Fig. \ref{Fig2}b. 

The Hamiltonian (\ref{eq2}) for the medial lattice is then written in a pictorial
representation \cite{Pollmann06c} as  

\begin{eqnarray}
H_{\rm{eff}} & = & g \sum_{\{\smallrech,\smallrecv\}} \Big( \big|\emptya
\big\rangle \big\langle\emptyb \big| - \big|\fulla \big\rangle
\big\langle\fullb \big|\Big)\nonumber \\
& = & g\sum_{\{\smallrech,\smallrecv\}} \Big(\big| A \big\rangle\big\langle
\overline{A} \big| - \big| B \big\rangle\big\langle \overline{B} \big| \Big)  
\label{eq3}
\end{eqnarray}

\noindent with $g = \frac{12 t^3}{V^2}$. The two ring-hopping processes differ
with respect to the site in the center of the flipable plaquette. It is
either empty (process A) or occupied (process B). For identification of
conserved quantities we divide the links of the square lattice into four
sublattices 1, ..., 4. It is seen immediately that H$_{\rm eff}$ conserves the
total number of particles (loop segments) on each sublattice $N_1, ...,
N_4$. Moreover $N_1 - N_3$ and $N_2 - N_4$ are also conserved. Note that the
latter two quantities are also conserved when higher order processes in $t/V$
are included while that is not the case for $N_1, ..., N_4$ separately. The
quantities $(N_1 - N_3)$ and $(N_2 - N_4)$ are related to the gradients
$\chi_x$ and $\chi_y$ of a height field $h(x,y)$ \cite{Bloete82}. A
height representation does apply here since the Hilbert space on which H$_{\rm
  eff}$ acts is equivalent to that of the six-vertex model
\cite{BaxterBook}.  

Returning to (\ref{eq3}) we note that the sign of $g$ is unimportant since it
can always be changed by a proper gauge transformation of the basis states
\cite{Pollmann06c}. The relative sign in (\ref{eq3}) results from a
difference in the occupation of the central site of the plaquette, when A and B
processes are accounted for. But as pointed out in Ref.~\cite{Boyarsky} also
  that sign can be changed by a special gauge transformation so that we do not
  face here the well-known fermionic sign problem. While processes of the form
  $\mid B \rangle \langle \bar{B} \mid$ in H$_{\rm eff}$ do not change the
  topology of a closed string (loop) configuration processes of the form $\mid
  A \rangle \langle \bar{A} \mid$ do so. As discussed in Ref.~\cite{Boyarsky} one loop can go over into three separate loops or two
  loops inside a third one and vice versa. Also two separate loops can go over
  into one loop inside a second one and vice versa. When the time evolution
  exp$\left[ -i\tau H_{\rm eff} \right]$ of the loops is considered one obtains
  world sheets instead of world lines. Because of the square lattice structure
  spatial changes are discretize. When the continuum limit is taken the loops
  or closed strings show a time evolution which has the form of world sheets
  and is schematically shown in Fig. \ref{Fig3}. While Fig. \ref{Fig3}a
  correspondents to B processes, Figs. \ref{Fig3}b-d result from A
  processes. The fluctuating loops depicted in that figure represent the ground
  state or the vacuum state of the system. It has been calculated before for
  finite systems by numerical methods and is well known
  \cite{Pollmann06b}.  

Next we consider the case when an energy $\Delta E$ is added to the
vacuum. When $\Delta E > V$ a loop is broken up and an open string is
created. One end of the string is touching a closed loop while the other end
is open. At those two points the tetrahedron rule is broken. In the original
checkerboard lattice they correspond to crisscrossed square with three
particles and with one particle only. Thus at the two ends of the string the
charges are $\frac{e}{2}$ and $-\frac{e}{2}$. We may speak also of a
particle-antiparticle pair production (see \cite{Cheng88}). A broken
loop on the medial lattice is shown in Fig. \ref{Fig4}a by the black line. The
time evolution of a configuration with loop breaking is shown in
Fig. \ref{Fig4}b when again the continuous limit is taken. 

It was demonstrated in \cite{Pollmann06b} that a particle pair with
charge $+\frac{e}{2}$, $-\frac{e}{2}$ is confined by a constant confining
force. This is reminiscent  of quarks. Thus a tension $T$ must be associated
with the string connecting the two fractions. The string tension was determined
to be $T = 0.2 \mid g \mid$. When a fractionally charged pair is pulled apart
so that the confining  energy exceeds the threshold value $\Delta E = V$, a new
particle-antiparticle pair with $\frac{e}{2}, -\frac{e}{2}$ is generated out of
the vacuum. This way the energy is lowered. Note the analogy to the generation
of $\mu$ mesons, i.e., quark-antiquark production in QCD
\cite{Cheng88}. Pumping an energy $\Delta E = nV$ into the 
system generates $n$ pairs with fractional charge $\frac{e}{2},
-\frac{e}{2}$. They can combine to form new 
pairs $\frac{e}{2}, \frac{e}{2}$ and $-\frac{e}{2}, -\frac{e}{2}$, depending on
relative distances of the constituents. Note that a string with charges
$+\frac{e}{2}, +\frac{e}{2}$ at its ends is touching two closed loops while a
string with charges $-\frac{e}{2}, -\frac{e}{2}$ at its end is not touching any
loops. For a better visualization this is shown in Fig. \ref{Fig5}a for two
pairs the medial square lattice. The corresponding time evolution of those two
pairs $\left( \frac{e}{2}, \frac{e}{2} \right)$ and $\left( -\frac{e}{2},
-\frac{e}{2} \right)$ is schematically indicated in Fig. \ref{Fig5}b.

It should be realized that one must distinguish between two types of pairs
$\left( \frac{e}{2}, \frac{e}{2} \right)$ and likewise $\left( -\frac{e}{2},
-\frac{e}{2} \right)$. One type is equivalent to adding or removing a spinless
particle. In that case the fractional charges must sit on different sublattices
of the checkerboard lattice. Adding a particle corresponds to converting an
empty link into an occupied one which connects two loops (see
Fig. \ref{Fig2}b). Removing a particle implies taking out a loop segment. In
both cases the two lattice sites connected by 
the link belong to different sublattices. We attach therefore an additional
colour index to the fractional particles, e.g., black or white depending on
which sublattice of the checkerboard lattice or the medial square lattice they
are situated. Pairs corresponding to electrons or holes are colour
neutral. They have equal number of black and white fractional parts. This is
the case for the pairs shown in Fig. \ref{Fig5}. When pairs $\left(
\frac{e}{2}, \frac{e}{2} \right)$ or $\left( -\frac{e}{2}, -\frac{e}{2}
\right)$ are not colour neutral they can not reduce to an electron or
hole. Instead they have always at least one string between the two
constituents. Their energy is higher than the one of an electron or hole.

When due to large values of $\Delta E$ the density of broken loops and hence
the generation of fractionally charged particle-antiparticle pairs is
sufficiently high, we are ending up with a plasma consisting of particles and
antiparticles with fractional charges. Although they are still connected
pairwise by strings which act like glue (i.e., gluons), the changes of
connections take place so frequently, namely with $|t|$ that the strings must
be also considered as part of the plasma. 

As $\Delta E$ becomes very large as compared with $V$ the significance of the
tetrahedron rule decreases. Not only will there be more and more tetrahedra
generated with three particles or one particle but also with four or
zero particles. Thus a string description looses its meaning. Instead it
is more appropriate to return to a description of the system in terms of
electrons with their correlation hole. Since the original loops are completely
disrupted the kinetic processes take place on the scale of $|t|$ rather than
$|t|^3/V^2$.

\section{Other frustrated lattices}

\label{Sect:OthFrusLatt}

\begin{figure}[thb]
 \begin{center}
      \begin{tabular}{cc}
	(a)~\includegraphics[width=3cm]{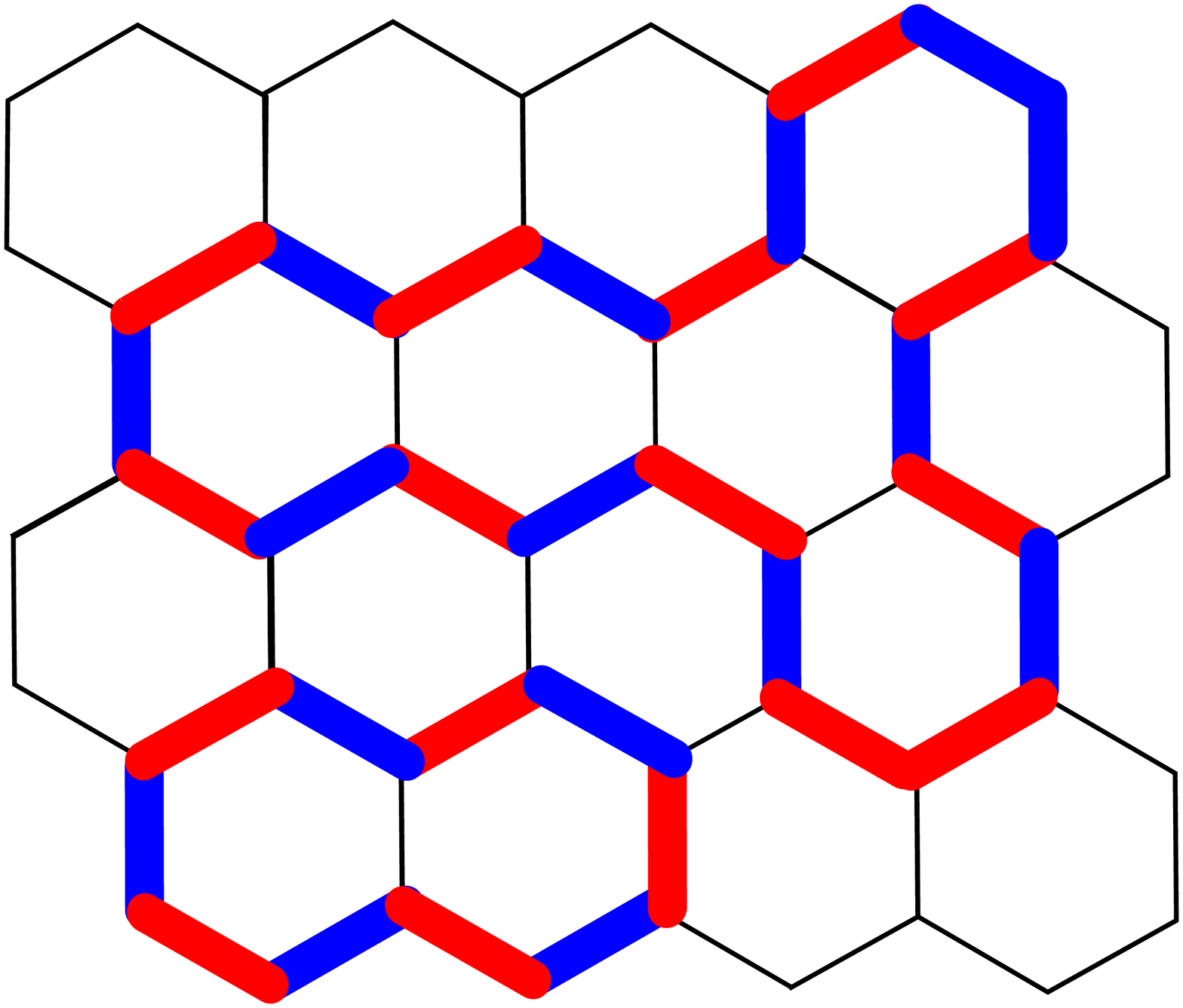}&
	(b)~\includegraphics[width=3cm]{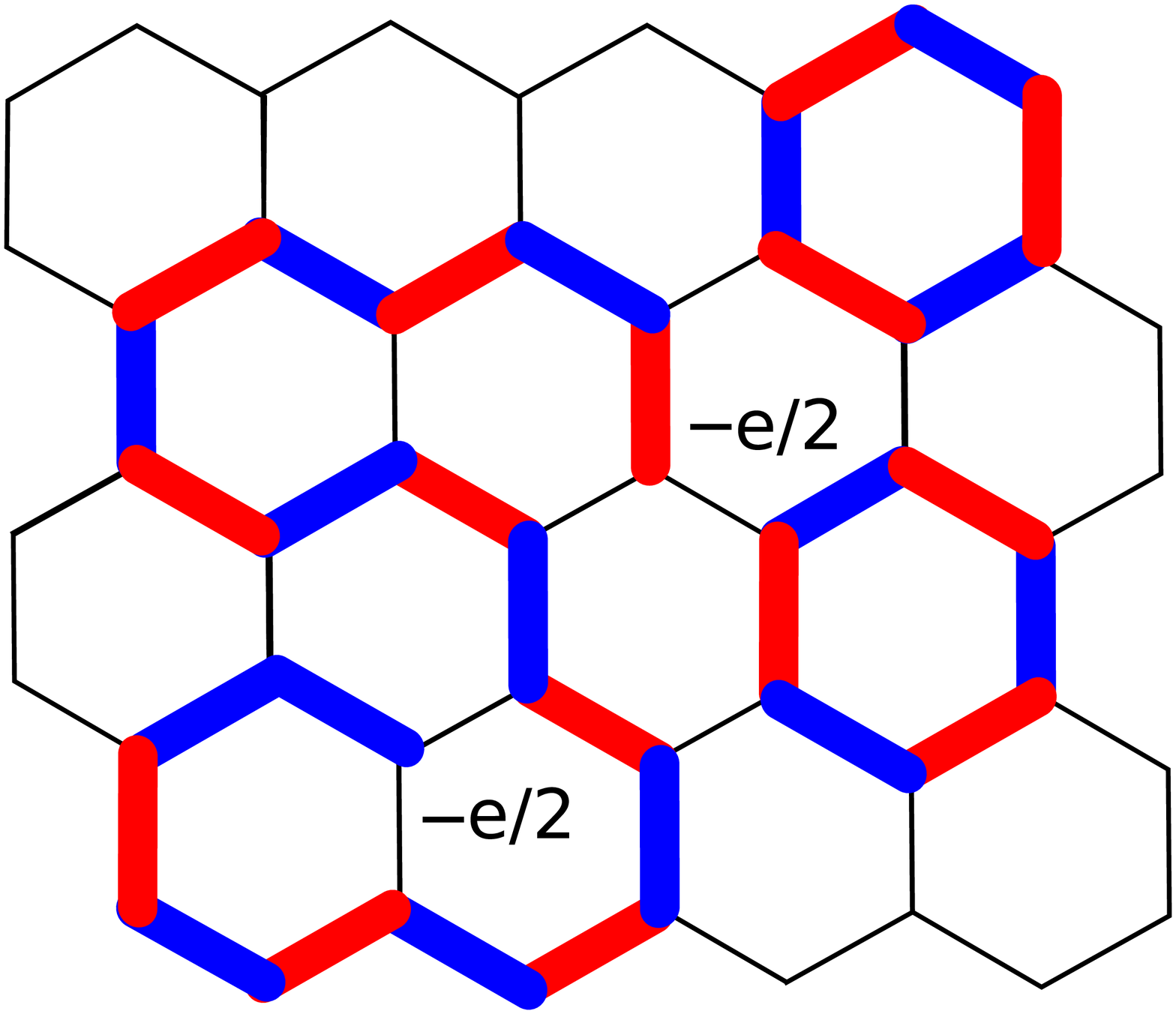}
      \end{tabular}
    \end{center}
    \caption{(a): Honeycomb lattice at 1/3 filling with electrons sitting on
	  coloured links, with red/blue referring to spin up/down.
(b): Removal of an electron with spin up resulting in an open string with
	  charges -e/2 at each end and a net down spin. This spin is highly
	  nonlocal and distributed over the open string. The latter is a Heisenberg
	  chain with $S_z = -1/2$. Figs (a) and (b) differ in addition by a hexagon flipping process.}
\label{Fig7}
\end{figure}
~

There exist many geometrically frustrated lattice systems. An instructive
overview is found in \cite{Diep05}. Two particularly well investigated
lattice types 
are the pyrochlore and the kagom\`e lattice. As pointed out before the
checkerboard lattice and the pyrochlore lattice are closely related, one being
the two-dimensional version of the other. We consider spinless fermions on a
pyrochlore lattice at half filling described by the Hamiltonian (\ref{eq1}). In
the absence of hopping, i.e., for $t = 0$ the ground state is again macroscopical degenerate. As before we are interested in strongly correlated
systems in which case $V \gg |t|$. Like in a checkerboard lattice terms of
order $t^2/V$ do not lift the degeneracy of the ground state. But terms of
order $t^3/V^2$ do. Therefore an effective Hamiltonian of the form (\ref{eq2})
is again applicable for describing the low-energy dynamics of the
system. However when ring hopping takes place there is unlike in the
checkerboard lattice no site inside the ring. This is seen explicitely by going
over to the medial lattice which in this case has the diamond structure. It is
obtained by connecting the clusters of neighboring tetrahedra. An example of a
configuration with a flipable hexagon in the medial lattice is shown in
Fig. \ref{Fig6}. Thus processes of type $\mid B \rangle \langle \bar{B} \mid$
found in configurations of the checkerboard lattice are ruled out here. This
implies that all fluctuations are accompanied by topological changes in the
loop covering. When a particle of charge $e$ is added to the system an
unoccupied link changes into an occupied one and the two associated sites have
three occupied links ending at them. By hopping processes of order 
$t$ the two sites with three links separate carrying a charge $\frac{e}{2}$
each. Various authors \cite{Balents02,Hermele04} predict the existence
of a deconfined phase on related dimer/loop models on the diamond lattice. This
suggests that deconfined fractional charges on the pyrochlore lattice can exist
as conjectured in \cite{FuldeP02}. The processes determining the time
evolution of the strings are the same as in Figs. \ref{Fig3}b-d. 

Of interest is also the case of $\frac{1}{4}$ filling. The strong repulsions
require that each tetrahedron contains one particle. Its time evolution is
governed again by (\ref{eq2}). The associated world lines of the particles are
subject to strong nonlocal subsidiary conditions, i.e., that every tetrahedron
contains exactly one particle. Their propagation is equivalent to that of a
string net. This is seen by connecting all empty sites and studying their time
evolution as that of a net of strings. 

Next we consider a kagom\`e lattice at $\frac{1}{3}$ filling with electrons
obeying an extended Hubbard Hamiltonian

\begin{equation}
H = -t \sum_{\langle ij \rangle\sigma} \left( c^+_{i \sigma} c_{j \sigma} +
h.c. \right) + V \sum_{\langle ij \rangle} n_i n_j + U \sum_i n_{i \uparrow}
n_{i \downarrow} ~~~.  
\label{eq4}
\end{equation} 

\noindent We assume $U \rightarrow \infty$ so that double occupancies of sites
are excluded and furthermore that $V \gg \mid t \mid$, i.e., that the system is
in the strong coupling limit. We eliminate $U$ from the Hamiltonian by
replacing $H$ by an extended $t-J$ Hamiltonian \cite{Brooks} of the
form 

\begin{equation}
\tilde{H} = -t \sum_{\langle ij \rangle} \left( \hat{c}^+_{i \sigma} c_{j
  \sigma} + h.c. \right) + V \sum_{\langle ij \rangle} n_i n_j + J
  \sum_{\langle ij \rangle} \left( {\bf S}_i {\bf S}_j - \hat{n}_i \hat{n}_j
  \right) ~~~.   
\label{eq5}
\end{equation} 

Here
\begin{eqnarray}
\hat{c}^+_{i \sigma} & = & c^+_{i \sigma} \left( 1 - n_{i - \sigma}
\right)\nonumber \\ 
\hat{c}_{i \sigma} & = & c_{i \sigma} \left( 1 - n_{i - \sigma} \right)
\label{eq6}
\end{eqnarray} 

\noindent act in the reduced Hilbert space, i.e., one which does not contain
doubly occupied sites. Furthermore $\hat{n}_i = \sum_\sigma \hat{c}^+_{i
  \sigma} \hat{c}_{i \sigma}$ and ${\bf S}_i = (1/2) \sum_{\alpha \beta}
\hat{c}^+_{i \alpha} {\bf \sigma}_{\alpha \beta} \hat{c}_{i \beta}$. The
$\bf\sigma$ matrices are the usual Pauli matrices. The spin coupling constant
is $J = \frac{4t^2}{U}$. The large on-site repulsion U has been replaced by an
effective antiferromagnetic spin-spin interaction which accounts for virtual
hopping processes by which a site is doubly occupied for a short time. A
transformation from (\ref{eq4}) to (\ref{eq5}) leads to additional terms of
order $t^2/U$ involving three sites, but they are usually neglected. Clearly $V
\gg J$.

The new aspect here is the inclusion of spin degrees of freedom. When $t = 0$
the ground state is again macroscopically degenerate. Going over to the medial
lattice which is the honeycomb one we connect the occupied links which leads to
a loop covering of the plane. The strings forming the loops consist of segments
of two different colours in correspondence to the spin of the electrons on the
links. This is shown in Fig. \ref{Fig7}a for a particular configuration. On
each loop the number of up and down spins is equal. The antiferromagnetic
interaction favors small loops since the energy per site of a Heisenberg chain
with an even number of sites increases with the chain length
\cite{Bloete86,Karbach95,Eggert}. When the dynamics is included we deal again
with ring hopping on hexagons of the form

\begin{equation}
H_{\rm{hex}}=-g\sum_{\left\{ \smallhexh\right\}
\left\{\blacktriangle\blacksquare\bullet\right\}}
\Big(\big|\hexaa\big\rangle\big\langle\hexb\big|
+\big|\hexab\big\rangle\big\langle\hexb\big|+{\rm H.c.}\Big)
\label{eq:Hhex}
\end{equation}

\noindent with $g=6t^{3}/V^{2}$. The sum over the three symbols is taken over all possible color (spin) combinations of a flippable hexagons. The particles can hop either clockwise or counter-clockwise around  hexagons. These processes can lead to different configurations, depending on the colors (spins) of the dimers on the hexagon. The evolution of loops with time is the same as the one due to $A$ processes of the checkerboard lattice and therefore of the form shown in Figs. \ref{Fig3}b-d. Note that we have here in addition fluctuations
of the spins of the loop segments. By writing

\begin{eqnarray}
J \sum_{\langle ij \rangle} \left( {\bf S}_i {\bf S}_j - \frac{\hat{n}_i
  \hat{n}_j}{4} \right) & = & J \sum_{\langle ij \rangle} \left( S^z_i
  S^z_j - \frac{\hat{n}_i \hat{n}_j}{4} \right)\nonumber \\
&& + \frac{J}{2} \sum_{\langle ij \rangle} \left( S^+_i S^-_j + S^-_i S^+_j
  \right) 
\label{eq8}
\end{eqnarray}

\noindent we notice that due to the last term on the right-hand side
neighboring links can change spins from up-down to down-up (see
Fig. \ref{Fig7}a). 

Next let us remove one electron from the system by eliminating one link in a
loop. This results in an open string. At the ends of that string charges
$-\frac{e}{2}$ are located. They can separate from each other due to hopping
processes. An example is shown in Fig. \ref{Fig7}b. The open ended chain
contains always an odd number of sites. The ground state of such a Heisenberg
chain is two-fold degenerate with total $S_z = \pm \frac{1}{2}$. Which of the
two possibilities realized depends on the spin of the removed electron. In any
case, it is noticed that the spin $\frac{1}{2}$ is now smeared (or distributed)
over the open string and is highly nonlocal. The time evolution of the open
string follows again the ones in Figs. \ref{Fig3}(b-d) but this time with one
opened loop at any time $t$.

In passing we note that it was shown in \cite{Pollmann07} that the
ground state of the system at $\frac{1}{6}$ filling is ferromagnetic with
maximum spin $S_{\rm tot} = \frac{1}{2} M$ where $M$ is the total particle
number. It was also shown that the ferromagnetic ground state is robust against
perturbations. Due to the inclusion of spin degrees of freedom there is now an
additional lower energy scale present which is of order $J$. Since we are
interested here in pointing out the connections of strongly correlated
electrons with strings only, we defer a detailed discussion of this scale to a
future investigation.

\section{Summary}

\label{Sect:Summary}

The aim of this paper has been to draw attention to the fact that in strongly
correlated electron systems on frustrated lattices the dynamics of the system
is given by that of closed and open strings. For simplicity we have assumed in
the larger part of the paper that the electrons are fully spin polarized or
spinless. That enabled us to focus on charge degrees of freedom. At particular
lattice filling factors, the ground state of the system consists of fluctuating
loops which cover completely the lattice. The strings are a direct consequence
of strong nonlocal subsidiary conditions. In the case of a pyrochlore or
checkerboard lattice they take the form of a tetrahedron rule which ensures
that nearest-neighbor repulsions of fully polarized electrons remain as small
as possible. The time evolution of strings results in world sheets instead of
world lines. The dynamics caused by the kinetic energy reduces in the
strong-coupling limit considered here to ring hopping processes. This assumes
special filling factors like the one mainly considered here of one half.

Dynamical fluctuation lead not only to deformations of loops but
aölso to loop formations of different topologies. In the continuum limit, which
we consider here, the discrete lattice structure is ignored and enters only in
form of the dynamic processes which depend on it. The different topological
changes in the loop covering induced by the dynamics have been pointed out and
visualized. 

When energy is added to such a system which exceeds a threshold value, closed
strings, i.e., loops begin to break up. In that case an open string has at its
ends two particles with charges $+\frac{e}{2}$ and $-\frac{e}{2}$. This
corresponds to half of an electron and half of a hole with respect to the
vacuum, i.e., the ground state. The string connecting the two has a string
tension which is finite in the two-dimensional checkerboard case but is
expected to be zero in the three dimensional pyrochlore lattice. With
increasing energy the number of open strings increases and a plasma of
fractionalized particles form until the string picture looses its meaning. 

When the electron spin is taken into consideration strings are coloured
according to the spins of the involved particles. This has been explicitely
discussed for a kagom\'e lattice at $\frac{1}{3}$ filling. Removal of an
electron leads to an open string with a spin $\frac{1}{2}$ distributed all over
the string. The same holds true when the pyrochlore lattice is considered
instead. We are dealing here with a three dimensional system with spin-charge
separation. This contradicts common belief that the latter can occur only in
low-dimensional systems.

As noticed above strongly correlated electrons on frustrated lattices
show features akin to particle physics. It would indeed be strange if nature
would realize certain phenomena only in one field of physics and not also in
others albeit in varied form.


\begin{thebibliography}{30}
\expandafter\ifx\csname natexlab\endcsname\relax\def\natexlab#1{#1}\fi
\expandafter\ifx\csname bibnamefont\endcsname\relax
  \def\bibnamefont#1{#1}\fi
\expandafter\ifx\csname bibfnamefont\endcsname\relax
  \def\bibfnamefont#1{#1}\fi
\expandafter\ifx\csname citenamefont\endcsname\relax
  \def\citenamefont#1{#1}\fi
\expandafter\ifx\csname url\endcsname\relax
  \def\url#1{\texttt{#1}}\fi
\expandafter\ifx\csname urlprefix\endcsname\relax\def\urlprefix{URL }\fi
\providecommand{\bibinfo}[2]{#2}
\providecommand{\eprint}[2][]{\url{#2}}

\bibitem{Kuzmany}
\bibinfo{editor}{\bibfnamefont{H.}~\bibnamefont{Kuzmany}},
  \bibinfo{editor}{\bibfnamefont{M.}~\bibnamefont{Mehring}}, \bibnamefont{and}
  \bibinfo{editor}{\bibfnamefont{J.}~\bibnamefont{Fink}}, eds.,
  \emph{\bibinfo{title}{Electronic Properties of High-T$_c$ Superconductors}},
  vol. \bibinfo{volume}{113} of \emph{\bibinfo{series}{Springer Series in Solid
  State Sciences}} (\bibinfo{publisher}{Springer, Heidelberg},
  \bibinfo{year}{1993}).

\bibitem{Bednorz}
\bibinfo{editor}{\bibfnamefont{J.~G.} \bibnamefont{Bednorz}} \bibnamefont{and}
  \bibinfo{editor}{\bibfnamefont{K.~A.} \bibnamefont{M\"uller}}, eds.,
  \emph{\bibinfo{title}{Superconductivity}}, vol.~\bibinfo{volume}{90} of
  \emph{\bibinfo{series}{Springer Series in Solid State Sciences}}
  (\bibinfo{publisher}{Springer, Heidelberg}, \bibinfo{year}{1990}).

\bibitem{Plakida95}
\bibinfo{author}{\bibfnamefont{N.~M.} \bibnamefont{Plakida}},
  \emph{\bibinfo{title}{High-Temperature Superconductivity}}
  (\bibinfo{publisher}{Springer}, \bibinfo{address}{Heidelberg},
  \bibinfo{year}{1995}).

\bibitem{Dagotto}
\bibinfo{author}{\bibfnamefont{E.}~\bibnamefont{Dagotto}},
  \emph{\bibinfo{title}{Nanoscale Phase Separation and Colossal
  Magnetoresistance: The Physics of Manganites and Related Compounds}}, vol.
  \bibinfo{volume}{136} of \emph{\bibinfo{series}{Springer Series in Solid
  State Sciences}} (\bibinfo{publisher}{Springer}, \bibinfo{address}{Berlin},
  \bibinfo{year}{2003}).

\bibitem{Imada98}
\bibinfo{author}{\bibfnamefont{M.}~\bibnamefont{Imada}},
  \bibinfo{author}{\bibfnamefont{A.}~\bibnamefont{Fujimori}}, \bibnamefont{and}
  \bibinfo{author}{\bibfnamefont{Y.}~\bibnamefont{Tokura}},
  \bibinfo{journal}{Rev. Mod. Phys.} \textbf{\bibinfo{volume}{70}},
  \bibinfo{pages}{1039} (\bibinfo{year}{1998}).

\bibitem{Kudasov03}
\bibinfo{author}{\bibfnamefont{Y.~B.} \bibnamefont{Kudasov}},
  \bibinfo{journal}{Physics-Uspekhi} \textbf{\bibinfo{volume}{46}},
  \bibinfo{pages}{117} (\bibinfo{year}{2003}).

\bibitem{Stewart}
\bibinfo{author}{\bibfnamefont{G.~R.} \bibnamefont{Stewart}},
  \bibinfo{journal}{Rev. Mod. Phys.} \textbf{\bibinfo{volume}{56}},
  \bibinfo{pages}{755} (\bibinfo{year}{1984}).

\bibitem{Norman}
\bibinfo{author}{\bibfnamefont{M.~R.} \bibnamefont{Norman}} \bibnamefont{and}
  \bibinfo{author}{\bibfnamefont{D.}~\bibnamefont{Koelling}},
  \emph{\bibinfo{title}{In: Handbook of the Physics and Chemistry of Rare
  Earth}}, vol.~\bibinfo{volume}{17} (\bibinfo{publisher}{Elsevier},
  \bibinfo{address}{Amstersam}, \bibinfo{year}{1993}).

\bibitem{ZwiebachBook}
\bibinfo{author}{\bibfnamefont{B.}~\bibnamefont{Zwiebach}},
  \emph{\bibinfo{title}{A first course in string theory}}
  (\bibinfo{publisher}{Cambridge Univ. Press}, \bibinfo{address}{Cambridge},
  \bibinfo{year}{2004}).

\bibitem{Boyarsky}
\bibinfo{author}{\bibfnamefont{A.}~\bibnamefont{Boyarsky}},
  \bibinfo{author}{\bibfnamefont{B.}~\bibnamefont{Kulik}}, \bibnamefont{and}
  \bibinfo{author}{\bibfnamefont{O.}~\bibnamefont{Ruchayskiy}}
  (\bibinfo{year}{2003}), \eprint{arXiv:hep-th/0312242}.

\bibitem{Bergmann}
\bibinfo{author}{\bibfnamefont{O.}~\bibnamefont{Bergmann}}
  (\bibinfo{year}{2004}), \eprint{arXiv:hep-th/0401106}.

\bibitem{levin2005}
\bibinfo{author}{\bibfnamefont{M.~A.} \bibnamefont{Levin}} \bibnamefont{and}
  \bibinfo{author}{\bibfnamefont{X.-G.} \bibnamefont{Wen}},
  \bibinfo{journal}{Phys. Rev. B} \textbf{\bibinfo{volume}{71}},
  \bibinfo{pages}{045110} (\bibinfo{year}{2005}).

\bibitem{Moessner04a}
\bibinfo{author}{\bibfnamefont{R.}~\bibnamefont{Moessner}},
  \bibinfo{author}{\bibfnamefont{O.}~\bibnamefont{Tchernyshyov}},
  \bibnamefont{and} \bibinfo{author}{\bibfnamefont{S.~L.}
  \bibnamefont{Sondhi}}, \bibinfo{journal}{J. Stat. Phys.}
  \textbf{\bibinfo{volume}{116}}, \bibinfo{pages}{755} (\bibinfo{year}{2002}).

\bibitem{Pollmann06c}
\bibinfo{author}{\bibfnamefont{F.}~\bibnamefont{Pollmann}},
  \bibinfo{author}{\bibfnamefont{J.}~\bibnamefont{Betouras}},
  \bibinfo{author}{\bibfnamefont{K.}~\bibnamefont{Shtengel}}, \bibnamefont{and}
  \bibinfo{author}{\bibfnamefont{P.}~\bibnamefont{Fulde}},
  \bibinfo{journal}{Phys. Rev. Lett} \textbf{\bibinfo{volume}{97}},
  \bibinfo{pages}{170407} (\bibinfo{year}{2006}).

\bibitem{Lieb67}
\bibinfo{author}{\bibfnamefont{L.~H.} \bibnamefont{Lieb}},
  \bibinfo{journal}{Phys. Rev. Lett.} \textbf{\bibinfo{volume}{18}},
  \bibinfo{pages}{692} (\bibinfo{year}{1967}).

\bibitem{Anderson56}
\bibinfo{author}{\bibfnamefont{P.~W.} \bibnamefont{Anderson}},
  \bibinfo{journal}{Phys. Rev.} \textbf{\bibinfo{volume}{102}},
  \bibinfo{pages}{1008} (\bibinfo{year}{1956}).

\bibitem{Runge04}
\bibinfo{author}{\bibfnamefont{E.}~\bibnamefont{Runge}} \bibnamefont{and}
  \bibinfo{author}{\bibfnamefont{P.}~\bibnamefont{Fulde}},
  \bibinfo{journal}{Phys. Rev. B} \textbf{\bibinfo{volume}{70}},
  \bibinfo{pages}{245113} (\bibinfo{year}{2004}).

\bibitem{Bloete82}
\bibinfo{author}{\bibfnamefont{H.~W.~J.} \bibnamefont{Bl\"ote}}
  \bibnamefont{and} \bibinfo{author}{\bibfnamefont{H.~J.}
  \bibnamefont{Hilhorst}}, \bibinfo{journal}{J. Phys. A}
  \textbf{\bibinfo{volume}{15}}, \bibinfo{pages}{L631} (\bibinfo{year}{1982}).

\bibitem{BaxterBook}
\bibinfo{author}{\bibfnamefont{R.}~\bibnamefont{Baxter}},
  \emph{\bibinfo{title}{Exactly Solved Models in Statistical Mechanics}}
  (\bibinfo{publisher}{Academic, San Diego}, \bibinfo{year}{1982}).

\bibitem{Pollmann06b}
\bibinfo{author}{\bibfnamefont{F.}~\bibnamefont{Pollmann}} \bibnamefont{and}
  \bibinfo{author}{\bibfnamefont{P.}~\bibnamefont{Fulde}},
  \bibinfo{journal}{Europhys. Lett.} \textbf{\bibinfo{volume}{75}},
  \bibinfo{pages}{133} (\bibinfo{year}{2006}).

\bibitem{Cheng88}
\bibinfo{author}{\bibfnamefont{T.-P.} \bibnamefont{Cheng}} \bibnamefont{and}
  \bibinfo{author}{\bibfnamefont{L.-F.} \bibnamefont{Li}},
  \emph{\bibinfo{title}{Gauge Theory of elementary particle physics}}
  (\bibinfo{publisher}{Oxford University Press, USA}, \bibinfo{year}{1988}).

\bibitem{Diep05}
\bibinfo{editor}{\bibfnamefont{H.~T.} \bibnamefont{Diep}}, ed.,
  \emph{\bibinfo{title}{Frustrated Spin Systems}} (\bibinfo{publisher}{World
  Scientific, Singapore}, \bibinfo{year}{2005}).

\bibitem{Balents02}
\bibinfo{author}{\bibfnamefont{L.}~\bibnamefont{Balents}},
  \bibinfo{author}{\bibfnamefont{M.~P.~A.} \bibnamefont{Fisher}},
  \bibnamefont{and} \bibinfo{author}{\bibfnamefont{S.~M.}
  \bibnamefont{Girvin}}, \bibinfo{journal}{Phys. Rev. B}
  \textbf{\bibinfo{volume}{65}}, \bibinfo{pages}{224412}
  (\bibinfo{year}{2002}).

\bibitem{Hermele04}
\bibinfo{author}{\bibfnamefont{M.}~\bibnamefont{Hermele}},
  \bibinfo{author}{\bibfnamefont{M.~P.~A.} \bibnamefont{Fisher}},
  \bibnamefont{and} \bibinfo{author}{\bibfnamefont{L.}~\bibnamefont{Balents}},
  \bibinfo{journal}{Phys. Rev. B} \textbf{\bibinfo{volume}{69}},
  \bibinfo{pages}{064404} (\bibinfo{year}{2004}).

\bibitem{FuldeP02}
\bibinfo{author}{\bibfnamefont{P.}~\bibnamefont{Fulde}},
  \bibinfo{author}{\bibfnamefont{K.}~\bibnamefont{Penc}}, \bibnamefont{and}
  \bibinfo{author}{\bibfnamefont{N.}~\bibnamefont{Shannon}},
  \bibinfo{journal}{Ann. Phys. (Leipzig)} \textbf{\bibinfo{volume}{11}},
  \bibinfo{pages}{892} (\bibinfo{year}{2002}).

\bibitem{Brooks}
\bibinfo{author}{\bibfnamefont{A.~B.} \bibnamefont{Harris}} \bibnamefont{and}
  \bibinfo{author}{\bibfnamefont{R.~V.} \bibnamefont{Lange}},
  \bibinfo{journal}{Phys. Rev.} \textbf{\bibinfo{volume}{157}},
  \bibinfo{pages}{295} (\bibinfo{year}{1967}).

\bibitem{Bloete86}
\bibinfo{author}{\bibfnamefont{H.~W.~J.} \bibnamefont{Bl\"ote}},
  \bibinfo{author}{\bibfnamefont{J.~L.} \bibnamefont{Cardy}}, \bibnamefont{and}
  \bibinfo{author}{\bibfnamefont{M.~P.} \bibnamefont{Nightingale}},
  \bibinfo{journal}{Phys. Rev. Lett.} \textbf{\bibinfo{volume}{56}},
  \bibinfo{pages}{742} (\bibinfo{year}{1986}).

\bibitem{Karbach95}
\bibinfo{author}{\bibfnamefont{M.}~\bibnamefont{Karbach}} \bibnamefont{and}
  \bibinfo{author}{\bibfnamefont{G.}~\bibnamefont{M\"uller}},
  \bibinfo{journal}{J. Phys. A: Math. Gen.} \textbf{\bibinfo{volume}{28}},
  \bibinfo{pages}{4469} (\bibinfo{year}{1995}).

\bibitem{Eggert}
\bibinfo{author}{\bibfnamefont{S.}~\bibnamefont{Eggert}},
  \bibinfo{howpublished}{Priv. comm.}

\bibitem{Pollmann07}
\bibinfo{author}{\bibfnamefont{F.}~\bibnamefont{Pollmann}},
  \bibinfo{author}{\bibfnamefont{K.}~\bibnamefont{Shtengel}}, \bibnamefont{and}
  \bibinfo{author}{\bibfnamefont{P.}~\bibnamefont{Fulde}}
  (\bibinfo{year}{2007}), \eprint{arXiv:0705.3941}.

\end{thebibliography}
\end{document}